\documentclass[%
 reprint,
 amsmath,amssymb,
 aps,
pra,
]{revtex4-1}

\usepackage{color}
\usepackage[table]{xcolor}
\usepackage{graphicx}% Include figure files
\usepackage{dcolumn}% Align table columns on decimal point
\usepackage{bm}% bold math
\definecolor{lightlightgrey}{gray}{0.92}

\newenvironment{exer*}
  {\ex}
  {\endex}

\def\be{\begin{equation}}
\def\ee{\end{equation}}
\def\bea{\begin{eqnarray}}
\def\eea{\end{eqnarray}}

\usepackage{xspace} % Used for printing a trailing space better than using a tilde (~) using the \xspace command

\newcommand{\hairsp}{\hspace{1pt}} % Command to print a very short space
\newcommand{\ie}{\textit{i.\hairsp{}e.}\xspace} % Command to print i.e.
\newcommand{\intdA}{\int\!\!\textrm{dA}}

\newcommand{\prens}[1]{\left(#1\right)}
\newcommand{\vecb}[1]{\mathbf{#1}}
\newcommand{\Pow}{\mathcal{P}}

\newcommand{\bhat}[1]{\vecb{\hat{#1}}}
    \newcommand{\zhat}{\bhat{z}}

\begin{document}

\preprint{APS/123-QED}

\title{Guided acoustic and optical waves in silicon-on-insulator for Brillouin scattering and optomechanics}
\author{Christopher J. Sarabalis}%
\email{sicamor@stanford.edu}

\author{Jeff T. Hill}

\author{Amir H. Safavi-Naeini}
\email{safavi@stanford.edu}

\affiliation{%
Department of Applied Physics, and Ginzton Laboratory, Stanford
University, Stanford, California 94305, USA
}%

\date{\today}% It is always \today, today,

\begin{abstract}
We numerically study silicon waveguides on silica showing that it is possible to simultaneously guide optical and acoustic waves in the technologically important silicon on insulator (SOI) material system. Thin waveguides, or fins, exhibit geometrically softened mechanical modes at gigahertz frequencies with phase velocities below the Rayleigh velocity in glass, eliminating acoustic radiation losses. We propose slot waveguides on glass with telecom optical frequencies and strong radiation pressure forces resulting in Brillouin gains on the order of 500 and 50,000 1/(Wm) for backward and forward Brillouin scattering, respectively.

\end{abstract}

\pacs{Valid PACS appear here}

\maketitle

%\tableofcontents

\section{Introduction}

Stimulated Brillouin Scattering (SBS) arises from the interaction of propagating acoustic and optical fields. 
In many materials including silicon, Brillouin scattering is the strongest optical nonlinearity~\cite{Chiao1964,Shen1965}.
These interactions are thereby an opportune mechanism for on-chip nonlinear optical devices compatible with a rapidly growing silicon photonics toolbox~\cite{Pant2011a,Lee2012,Li2014a,VanLaer2015,Rakich2012,Eggleton2013,Kittlaus2015}.

While silicon on insulator (SOI) is a natural platform for photonics,  it is not as well-suited for on-chip phononics and acoustic waveguiding.
The high index contrast between silicon and silica glass readily allows for confinement of optical fields to the silicon device layer.
But silicon's relative stiffness -- and therefore high sound velocity -- makes guiding acoustic waves challenging, motivating the use of soft chalcogenide glasses and partial or complete releases (removal of the underlying glass) to co-localize optics and mechanics.
Chalcogenide glasses have exhibited the largest Brillouin gains, but use of non-standard materials makes integration with silicon foundry processes challenging~\cite{Pant2014}.
Released and partially released silicon structures, while using standard materials, require highly sensitive post-fabrication processing steps. Furthermore, while release steps minimize mechanical coupling to the environment, they also isolate devices from heat-dissipating thermal linkages and on-chip electronics.

 The intuition of ``index-guided'' confinement of mechanical modes fails for sub-wavelength structures. 
 Just as a beam can be made compliant if made thin, thin silicon strips or fins on glass can be made to confine mechanics despite  bulksilica's low  sound velocity. 
 This is related to how Rayleigh waves are confined to a surface -- free space boundaries affect optics and mechanics in fundamentally different ways. For optics, light propagates faster in vacuum than in a homogeneous dielectric.  
In this sense, the vacuum is ``stiffer'' than the dielectric. 
The opposite is the case for acoustic waves; free space, or free boundaries, softens the structural response, decreasing the phase velocity of guided waves.

As we will show here, by creating a wavelength-scale guided-wave structure with carefully designed free boundaries, we can in principle confine both light and sound in \textit{unreleased} silicon on insulator waveguides. This confinement leads to  low loss acoustic waves as well as large optomechanical coupling.
In section~\ref{sec:modes} we present analysis and simulations showing the existence of low-loss mechanical modes confined to silicon waveguides on a silica slab.
In section~\ref{sec:interaction}, we present designs for structures with simultaneously bound, strongly interacting photonic and phononic modes.
These modes exhibit broadband backward Brillouin scattering without acoustic radiation into the substrate as well as low-loss, forward Brillouin scattering.  These results open the way for a new class of fully CMOS compatible silicon photonic devices that utilize mechanical degrees of freedom.

\begin{figure}
    \vspace{-5mm}
    \centering
    \includegraphics[width=2.25in]{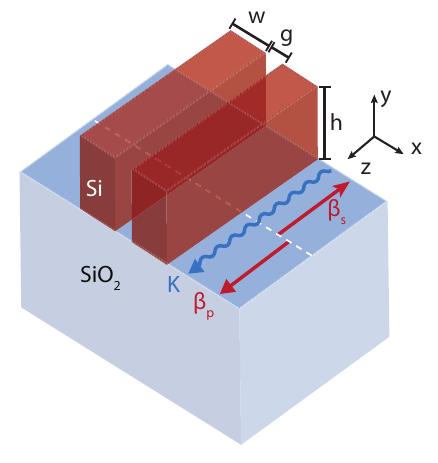}
    \caption{An SOI slot waveguide with height $h = 340~\textrm{nm}$, waveguide width $w = 100~\textrm{nm}$, and gap $g = 40~\textrm{nm}$ is optically single-moded in the telecom C-band and supports gigahertz frequency mechanical modes for on-chip backward and forward Brillouin scattering in a CMOS compatible architecture.}
    \label{fig:intro_waveguide}
    \vspace{-5mm}
\end{figure}

\section{Modal Description of Brillouin Scattering}

The interaction between propagating light and sound in a translationally invariant waveguide, as shown in Figure~\ref{fig:intro_waveguide}, is well described by the coupled mode theory of SBS.
The propagating electromagnetic and acoustic waves are described by vector fields
\begin{align}
    \label{eq:opt_modes}
    \vecb{E}\prens{x,y,z,t} &= \sum_i a_i(z,t)\vecb{e}_i\prens{x,y}e^{i\beta_i z - i\omega_i t}~~~\text{and} \\
    \label{eq:mech_modes}
    \vecb{U}\prens{x,y,z,t} &= b(z,t)\vecb{u}\prens{x,y}e^{\prens{i K - \frac{\alpha}{2}} z - i\Omega t},
\end{align}

respectively. 
These waves propagate along $\zhat$ as shown in Figure~\ref{fig:intro_waveguide}.
The envelope functions $a_i(z,t)$ and $b(z,t)$ capture slow variations of the amplitude of the modes such that $|\partial_z a_i(z,t)|~\ll~|\beta_i a_i(z,t)|$ and $|\partial_t a_i(z,t)|~\ll~|\omega_i a_i(z,t)|$. 

% Discuss phase matching. temporal
For there to be significant interactions both energy and momentum must be conserved.
Consider an incoming optical mode of frequency $\omega_\textrm{p}$ and a mechanical mode of frequency $\Omega$.
These modes can scatter into a downshifted optical \emph{Stokes wave} with frequency $\omega_\textrm{s} = \omega_\textrm{p} - \Omega$ or an upshifted \emph{anti-Stokes wave} with frequency $\omega_\textrm{p} + \Omega$.
Momentum conservation, or spatial phase-matching, for the Stokes process demands $\beta_\textrm{p} = \beta_\textrm{s}+ K$.
The process is commonly referred to as \emph{backward} and \emph{forward} SBS~\cite{Shelby1985} (BSBS and FSBS) in the case of counter- and co-propagating optical modes, respectively.
Additionally sound can scatter light between different modes of a waveguide, referred to as intermodal SBS or stimulated inter-polarization scattering (SIPS)~\cite{Russell1990}, which we do not consider here. 
These  conservation laws are reflected in the dynamics of the envelope functions as described by the coupled mode equations~\cite{Wolff2015}

\begin{align}
    \label{eq:cme1}   
       \prens{v_\textrm{p} \partial_z + \partial_t} a_\textrm{p} &= -i \tilde{g_0}^\ast a_\textrm{s} b, \\
    \label{eq:cme2}
           \prens{v_\textrm{s} \partial_z + \partial_t} a_\textrm{s} &= -i \tilde{g_0} a_\textrm{p} b^\ast, \\
    \label{eq:cme3}
           \prens{v_\textrm{b} \partial_z + \partial_t +  v_\textrm{b} (\alpha/2+ i \delta)} b &= -i \tilde{g_0} a_\textrm{p} a_\textrm{s}^\ast,
\end{align}
where $v_\textrm{p}$, $v_\textrm{s}$, and $v_\textrm{b}$ are the group velocities of the pump, stokes, and mechanical wave,  $\alpha$ is the attenuation constant for  acoustic waves, $\delta$ is the phase mismatch $K - (\beta_\textrm{p} -\beta_\textrm{s})$, and $\tilde{g_0}$ is the optomechanical coupling. 
These relations hold so long as the envelope functions vary slowly relative to $\beta$ and $\omega$ (or $K$ and $\Omega$ for mechanics).
Assuming the interaction is weak enough such that $a_\textrm{p} a_\textrm{s}^\ast$ is nearly constant over the mechanical attenuation length $\alpha^{-1}$, the mechanics mediates an effectively local interaction between the optical modes and Equations~(\ref{eq:cme1})-(\ref{eq:cme3}) can be solved for the steady-state dynamics of the optical fields:
\begin{align*}
    \partial_z|a_\textrm{p}|^2 &=  - 2\text{Re}\left\{ \frac{|\tilde{g_0}|^2}{v_\textrm{b} v_\textrm{p} (\alpha/2 +i\delta) } \right\} |a_\textrm{p}|^2|a_\textrm{s}|^2, \\
    \partial_z|a_\textrm{s}|^2 &=   2\text{Re}\left\{ \frac{|\tilde{g_0}|^2}{v_\textrm{b} v_\textrm{s} (\alpha/2 + i\delta)} \right\}  |a_\textrm{p}|^2|a_\textrm{s}|^2.
\end{align*}
The Brillouin gain parameter (2Re\{..\})  appearing in these equations depends on the normalization of the mode amplitudes. 
For the rest of this paper, we assume power normalization such that the calculated gain, $\Gamma$, has units $\text{W}^{-1}\text{m}^{-1}$ and is given by~\cite{Wolff2015}
\begin{equation}
    \Gamma = \frac{2\omega_{p}\Omega}{\Pow_\textrm{p} \Pow_\textrm{s} \Pow_\textrm{m}}\frac{\alpha/2}{\prens{\alpha/2}^2+\delta^2} \left| \intdA \;\vecb{e}^\ast_\textrm{s}\; \partial_u \varepsilon \; \vecb{u}\; \vecb{e}_\textrm{p}\right|^2,
    \label{eq:brillouinGain_BSBS}
\end{equation}
with modal powers $\Pow_{\{p,s,m\}}$.

\begin{figure*}
    \vspace{-5mm}
    \centering
    \includegraphics[width=\linewidth]{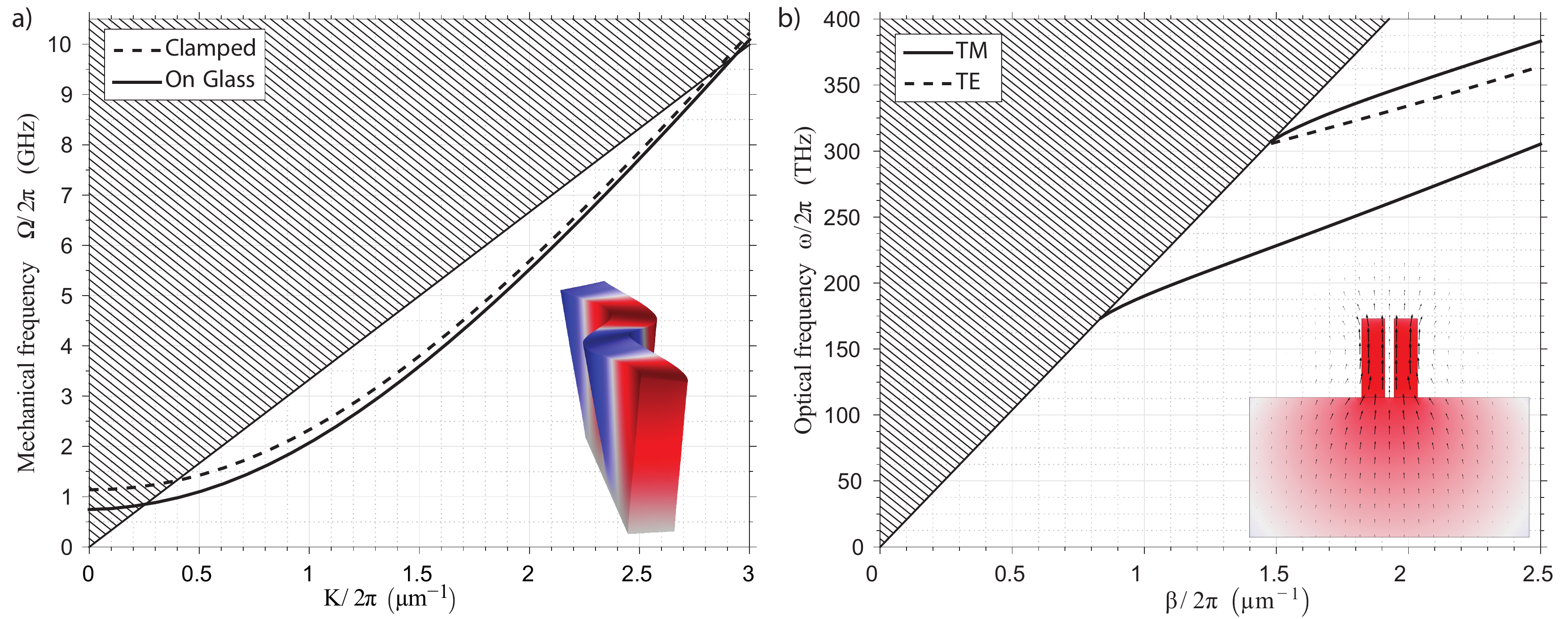}
    \caption{\textbf{a.} Mechanical bands of an $h = 340~\textrm{nm}, w = 100~\textrm{nm}$ strip waveguide for clamped (dashed) and on-glass (solid) models contain intervals of radiation-free modes falling below the Rayleigh cone (hatched). Supermodes of a slot waveguide (see Figure~\ref{fig:intro_waveguide}) are nearly degenerate with the on-glass strip waveguide band.
    \textbf{b.} Optical bands of a $\prens{h = 340, w = 100, g = 40}~\textrm{nm}$ slot waveguide show a single TM-like mode (solid) in the telecom C-band. 
    The fundamental TE-like band (dashed) emerges at higher frequencies due to the high aspect ratios necessary for the confinement of mechanics.}
    \label{fig:modes}
        \vspace{-5mm}
\end{figure*}

\section{Confining Light and Sound to Silicon on Glass}
\label{sec:modes}

Silicon's transverse sound speed $c_{\textrm{t},\textrm{Si}} = \sqrt{E/2(1+\nu)\rho} = 5390~\textrm{m/s}$ is much higher than in glass where $c_{\textrm{t},\textrm{g}}=3690~\textrm{m/s}$.
The material properties $E$, $\rho$, and $\nu$ are Young's modulus, density, and Poisson's ratio, respectively.
%
%  Mini Appendix on silicon mechanical properties.
%
% ( E [GPa] , rho [kg/m^3], nu )
%
% poly si   : (  169, 2330, 0.22)  [1 from 2]
% poly sio2 : (69-92, 2200, 0.17)  [1 from 3]
%
% Stiffness coefficients for crystalline silicon.
%
% (c_11 = 165 GPa, c_12 = 64 GPa, c_44 = 79.2 GPa) [1 from 4]
%
% [1] Cleland. Foundations of nanomechanics.
% [2] Sharpe et al.. New test structures and techniques for measurement of mechanical properties of MEMS materials. 1996
% [3] MT Kim. Influence of substrates on the elastic reaction of films for the microindentation tests. 1996.
% [4] Landolt-Bornstein
%
Vibrations of a silicon waveguide with phase velocities above $c_{\textrm{t},\textrm{g}}$ will readily excite surface and bulk waves of the glass substrate. Despite a relatively high material stiffness, high aspect-ratio silicon structures, as in Figure~\ref{fig:modes}, can have \emph{geometrically softened} modes with lower phase velocities than bulk and surface acoustic waves in glass.
We focus on strip waveguides in 340~nm thick SOI. Consider the vibrations of a 100~nm wide strip waveguide.
To simplify the analysis we first replace the glass substrate with fixed boundary conditions.
Solutions to the equations of elasticity are computed numerically in COMSOL~\cite{COMSOL5} for each wavevector $K$ by imposing Floquet boundary conditions on faces perpendicular to the axis of propagation.
A representative mechanical mode and the dispersion of the clamped beam is shown (dashed) in Figure~\ref{fig:modes}a.
At $K=0$ solutions reduce to the those of a cantilever for which \cite{Cleland2003}
\begin{equation}
    \left.\Omega\right|_{K = 0} = \sqrt{ \frac{EI}{\rho A} }\left(\frac{1.875}{h}\right)^2.
\end{equation}
Here $I = hw^3/12$ is the bending moment of the beam, $h$ and $w$ are the beam height and width, and $A$ the area of the profile.
We model silicon as an isotropic material with properties $E= 165~\textrm{GPa}$, $\rho = 2330~\textrm{kg/m}^3$, and $\nu = 0.22$~\cite{Cleland2003}.
As $K$ is increased the rectangular waveguide exhibits quadratic dispersion which is sensitive to $h$ and $w$.
Waveguide geometry becomes insignificant in the high-$K$ limit where the mechanical wavelength is small compared to $w$ and $h$.
In this limit the phase velocity limits to $c_{\textrm{t},\textrm{Si}}$ governed only by bulk material properties.

We now consider strip waveguides on glass.
The spectrum for this structure, as shown in Figure~\ref{fig:modes}a, is comprised of flexural modes of the beam (the solid line falling below the dashed clamped beam disperison discussed above), and surface waves at the glass-air interface (the hatched region) as well as bulk waves in the glass (also within the hatched region).
Notably for $w = 100~\textrm{nm}$, a broad interval of the flexural band lies below the dispersion of the surface acoustic waves described by the \emph{Rayleigh cone} with phase velocity $c_\text{R}$.

Scattering from this interval into the Rayleigh waves is forbidden by phase-matching considerations.
Here the Rayleigh wave velocity $c_\textrm{R}$ is related to the material properties of glass by the Rayleigh-Lamb equations.
Modeling the glass with material parameters $(E = 70~\textrm{GPa}, \rho = 2200~\textrm{kg}/\textrm{m}^3, \nu = 0.17)$, we find $c_\textrm{R} = 3333~\textrm{m/s}$~\cite{Cleland2003}.
The Rayleigh waves travel slower than longitudinal ($5848~\textrm{m/s}$) and transverse ($3690~\text{m/s}$) waves in glass.
The bulk waves in glass, also in the hatched region, lie above the Rayleigh line. 

As the width of a rectangular waveguide is increased, the flexural band rises in frequency.
At $w~=~250~\textrm{nm}$, the band lies completely above the Rayleigh line and no protected region is present.
Owing to the scale invariance of the equations of elasticity and the dispersion of acoustic waves, this characteristic aspect ratio 250:340 depends only upon material properties of the device and substrate layer and is therefore relevant to unreleased SOI devices of any thickness. For example, for 220 nm SOI, waveguides of width smaller than 160 nm are required to achieve non-leaky or phase-protected propagation.

\vspace{-3mm}
\subsection{Symmetry and the Optics of Slot Waveguides}
\label{sec:symmetryAndOptics}

The symmetry of strip waveguides leads to symmetries of the mode profiles, suppressing any intermodal Brillouin scattering via the acoustic waves studied above.
Under reflection through the $yz$ symmetry plane,  $\Theta_x$, the mechanical modes are either symmetric ($+$) or antisymmetric ($-$), \ie, $\vecb{U}\prens{\Theta_x\vecb{r}} = \pm\Theta_x\vecb{U}\prens{\vecb{r}}$.
The interaction rate in Equation~(\ref{eq:brillouinGain_BSBS}) is linear in the displacement field and both optical fields.
For intramodal scattering, since both optical modes are of the same symmetry, the interaction is nonzero only if the mechanical motion, $\vecb{U}\prens{x,y,z,t}$, is symmetric under $\Theta_x$.
As the flexural modes shown in Figure~\ref{fig:modes}a are antisymmetric, they cannot mediate intramodal scattering. In contrast, the symmetric mechanical supermode of side-by-side strip waveguides, \ie, a slot waveguide, can mediate intramodal scattering.
Tight confinement of the strip waveguide flexural modes inside the silicon causes the symmetric and antisymmetric supermodes of the slot structure to be nearly degenerate even at spacings below 100 nm, since very little acoustic coupling occurs through the glass. The splitting is therefore negligibly small on the scale of Figure~\ref{fig:modes}a, and is not visible.

The slot waveguide is a popular photonic structure.
The TE-like modes of slot waveguides are commonly used for sensing as electric field amplitudes are large in the gap~\cite{Almeida2004}. 
A recent study of SBS in chalcogenide glasses filling a slot waveguide has utilized this electric field enhancement to achieve large interactions with index-guided mechanical modes~~\cite{Mirnaziry2016}.
Slot waveguides have also been proposed for SBS in suspended silicon structures~\cite{VanLaer2014}.
Furthermore slotted structures have been implemented in photonic crystal fibers~\cite{Butsch2012} to achieve large optomechanical coupling. Consider the optical guided modes for $100~\textrm{nm}$ wide strips spaced $40~\textrm{nm}$ apart in $340~\textrm{nm}$ thick silicon shown in Figure~\ref{fig:modes}b.
For these dimensions, the waveguide has a single, TM-like mode in the telecom C-band.
The large splitting of the fundamental TE-like and TM-like modes of this waveguide is a consequence of the high aspect ratio necessary to confine mechanics. Additionally, we expect that by using TM guided modes, lower scattering  due to smaller fields at etched surfaces will lead to lower optical losses.
%Widening the waveguides and shrinking the slot width shifts the TE-like bands down faster than the TM-like bands. 

\begin{figure}
    \vspace{-5mm}
    \centering
    \includegraphics[width=0.9\linewidth]{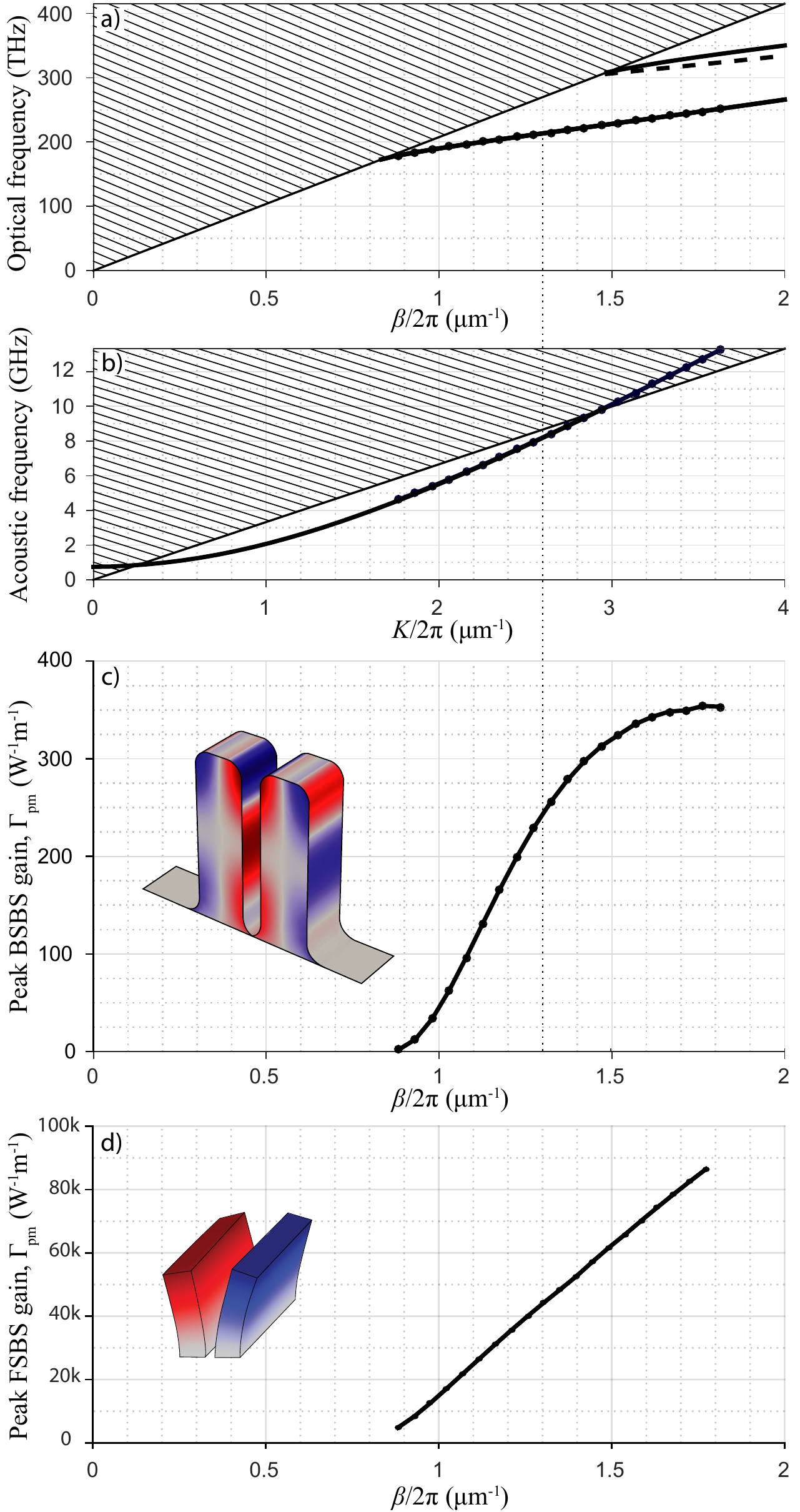}
    \caption{For BSBS, resontant optical and mechanical modes in \textbf{a.} and \textbf{b.} are aligned with the corresponding peak, or phase-matched, Brillouin gain $\Gamma_\textrm{pm}$ (\textbf{c.}). 
    The radiation pressure (boundary) contribution to the optomechanical coupling is an order of magnitude larger than the photoelastic effect (body), plotted inset to \textbf{c.} on separate scales for clarity.
    \textbf{d.} The peak FSBS gain is shown alongside the 873~MHz, $K=0$ mechanical mode. FSBS peak gains are $\sim200\times$ greater than BSBS gains, largely a consequence of lower mechanical frequency. We assume a mechanical quality factor of $Q = 1000$ in all calculations.}
    \label{fig:interaction}
        \vspace{-5mm}
\end{figure}

\vspace{-3mm}
\section{Backward and Forward Brillouin Scattering in Slot Waveguides}
\label{sec:interaction}

%We investigate the achievable Brillouin gains and the effects of the design parameters $w$ and $g$ in the slot geometry.

Having defined the acoustic and optical modes of interest, we now calculate the achievable Brillouin gains using Equation~(\ref{eq:brillouinGain_BSBS}). 
The displacement field causes a perturbation in the permittivity $\partial_u\varepsilon$ with a bulk contribution from the photoelastic effect $\partial_u\varepsilon_\textrm{pe}$ and a boundary contribution from radiation pressure $\partial_u\varepsilon_\textrm{rp}$.
For isotropic materials like silicon and glass, $\partial_u\varepsilon_\textrm{u} = -\varepsilon^2 \vecb{p} \vecb{S}$. 
The rank-2 photoelastic tensor $\vecb{p}$ relates the strain $S_{ij} = \frac{1}{2}\prens{\partial_i u_j + \partial_j u_i}$ to a change in $\varepsilon$~\cite{Yariv1984b}.  % could use citations to JChan thesis, maybe Yariv  (references in photoelastic code)
This term is integrated over the bulk of each dielectric.
A second contribution due to motion of the boundaries of the dielectrics takes the form of an integral over these surfaces with integrand~\cite{Johnson2002}
%\bea
%\vecb{e}_\textrm{s}^\ast\partial_u\varepsilon_\textrm{rp}\vecb{e}_\textrm{p} &=&\vecb{e}_{\textrm{s}\bot}^\ast\Delta\varepsilon\prens{\bhat{n}\cdot\vecb{u}}\vecb{e}_{\textrm{p}\bot} - \nonumber \\ && \vecb{d}_{\textrm{s}\parallel}^\ast\Delta\varepsilon^{-1}\prens{\bhat{n}\cdot\vecb{u}}\vecb{d}_{\textrm{p}\parallel}
%\eea
\begin{equation}
\vecb{e}_\textrm{s}^\ast\;\partial_u\varepsilon_\textrm{rp}\;\vecb{e}_\textrm{p} = 
\prens{\Delta\varepsilon\;\left|\vecb{e}_{\textrm{p}\bot}\right|^2 -    \Delta\varepsilon^{-1}\left|\vecb{d}_{\textrm{p}\parallel}\right|^2}{\bhat{n}\cdot\vecb{u}}   
\end{equation}
with the electric field perpendicular $\vecb{e}_\bot$ and electric displacement field parallel $\vecb{d}_\parallel$ to the normal of each boundary.
The permittivity differences $\Delta\varepsilon$ and $\Delta\varepsilon^{-1}$ are taken from outside to inside in the direction of the normal $\bhat{n}$.
Although in general both effects are significant in wavelength-scale waveguides~\cite{Rakich2010}, radiation pressure at the silicon-air boundary is an order of magnitude greater than all other bulk and boundary contributions for the slot waveguides considered and the values of $\Gamma$ calculated here.

\begin{figure}[h]
    \vspace{-2mm}
    \includegraphics[width=1.0\linewidth]{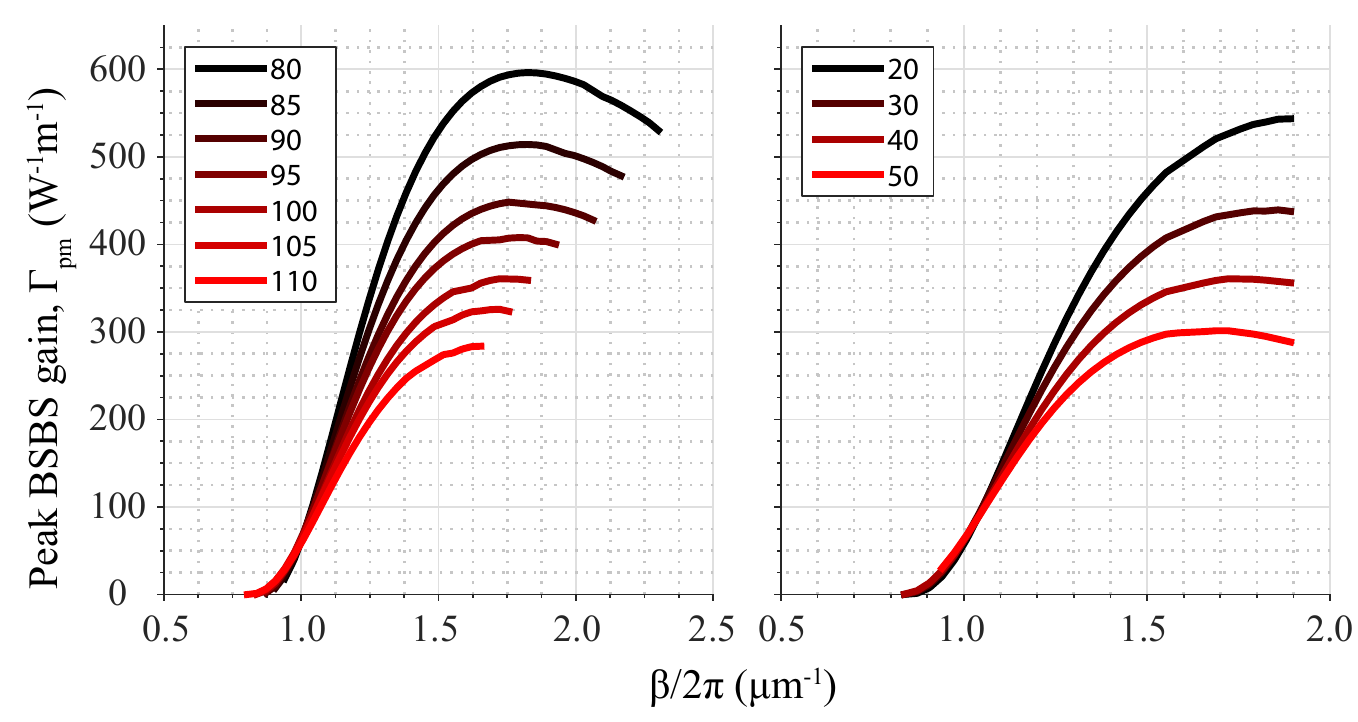}
    \caption{Phase-matched Brillouin gain for BSBS for different waveguide widths $w$ (labeled in nanometers) for fixed gap $g = 40~\textrm{nm}$, left, and different gaps $g$ (labeled in nanometers) for fixed width $w = 100~\textrm{nm}$, right.}
    \label{fig:geom_sweep}
        \vspace{-2mm}
\end{figure}

As the detuning $\delta = K - 2\beta$ in Equation~(\ref{eq:brillouinGain_BSBS}) is varied, $\Gamma$ sweeps out a Lorentzian with FWHM $\alpha$ and height equal to the \emph{peak}, or \emph{phase-matched}, \emph{Brillouin gain} $\Gamma_\textrm{pm} = \Gamma\big|_{\delta \small= 0}$.
For each optical wave vector $\beta$ (Fig.~\ref{fig:interaction}a) and corresponding resonant mechanical wave $K = -2\beta$ (Fig.~\ref{fig:interaction}b), we compute the phase-matched Brillouin gain for BSBS (Fig.~\ref{fig:interaction}c).
This gain $\Gamma_\textrm{pm}\prens{\beta}$ increases steadily from zero at $\beta_\textrm{min}$ before turning over approximately where the flexural mode enters the Rayleigh cone. In addition, we sweep the geometry and see in Figure~\ref{fig:geom_sweep}  that $\Gamma_\textrm{pm}$ increases with decreasing $w$ and $g$.

As discussed in Section~\ref{sec:symmetryAndOptics}, the antisymmetry of the flexural modes of a rectangular waveguide motivates the use of symmetric supermodes of the slot waveguide for intramodal SBS.
In Figure\xspace\ref{fig:interaction}c, a plot of the photoelastic and radiation pressure distributions show the drawbacks of the inherent asymmetry of the flexural modes.
These distributions are roughly antisymmetric about the $yz$ symmetry plane of each waveguide, cancelling in an optical field of constant intensity. Despite these cancellations, we obtain peak BSBS gains approaching 600~$\text{W}^{-1}\text{m}^{-1}$ that are comparable to silicon-based demonstrations.

% FSBS

For BSBS a large interval of the mechanical band is forbidden from radiating into the glass by phase-matching considerations.
In contrast, the $K=0$ mechanical mode radiates into surface acoustic waves and bulk acoustic modes of the glass.
Due to impedance mismatch, these loss channels do not necessarily prohibit the use of slot waveguides for FSBS.
Simulations of a $w = 100~\textrm{nm}$, $g = 40~\textrm{nm}$ slot waveguide show that acoustic radiation limits the quality factor to $Q = 1460$.
The 873~MHz mechanical resonance at $K=0$ radiates into glass modes at a rate of $ 600~\textrm{kHz}$.  % (see plots in soeye/slotwg/20160216_gammapoint)  
Current demonstrations of released, on-chip systems exhibit mechanical quality factors below this radiative limit \cite{VanLaer2015, Kittlaus2015}.  % VanLaer = 250;  max(Kittlaus) = 1020 \pm 50;

Since the modal power is not well defined around $K=0$, we express Equation\xspace(\ref{eq:brillouinGain_BSBS}) for FSBS on resonance as
\begin{equation}
    \Gamma_\textrm{pm} = \frac{2 \omega Q}{\Pow_\textrm{p} \Pow_\textrm{s} \mathcal{E}_\textrm{m}} \left| \intdA \;\vecb{e}^\ast_\textrm{s}\; \partial_u \varepsilon \; \vecb{u}\; \vecb{e}_\textrm{p}\right|^2,
    \label{eq:brillouinGain_FSBS}
\end{equation}
and use this to compute the $\Gamma_\textrm{pm}\prens{\beta}$ curve in Figure~\ref{fig:interaction}d.
The resulting peak Brillouin gain is two orders of magnitude larger than corresponding values for BSBS in agreement with the $\Omega^{-2}$ scaling of $\Gamma$ from the mechanical mode energy normalization in Equation~(\ref{eq:brillouinGain_FSBS}).
\vspace{-3mm}
\section{Conclusions}
\vspace{-2mm}

We have shown that it is possible to make waveguides that simultaneously guide interacting acoustic and optical waves in the technologically important silicon on insulator platform and presented a design for optically single-moded, Brillouin active waveguides in 340~nm SOI. These results open a significant new space of design freedom -- the incorporation of nanomechanics and acoustics in CMOS compatible photonic architectures, which can have applications in optical, RF-microwave, as well as novel sensing circuits.

\noindent\textbf{Acknowledgements} This work was supported by NSF ECCS-1509107 and the Stanford Terman Fellowship, as well as start-up funds from Stanford University. We thank Raphael van Laer for useful discussions.

%\renewcommand{\refname}{}

%\bibliographystyle{apsrev_ASN}
%\bibliography{Papers-SBS_1}
%\bibliography{LINQS-SBS_Papers}

\end{document}